\documentclass[a4paper,11pt]{article}
\usepackage{pos}
\usepackage{ulem}

\title{Left-hand cut and the HAL QCD method}

\author*[a]{Sinya Aoki}
\author[b]{Takumi Doi}
\author[b]{Yan Lyu}

\affiliation[a]{Center for Gravitational Physics and Quantum Information, Yukawa Institute for Theoretical Physics, Kyoto University,
  Kyoto 606-8502, Japan}

\affiliation[b]{Interdisciplinary Theoretical and Mathematical Physics Program (iTHEMS), RIKEN, Wako 351-0198, Japan}

\emailAdd{saoki@yukawa.kyoto-u.ac.jp}
\emailAdd{doi@ribf.riken.jp}
\emailAdd{yan.lyu@riken.jp}

\abstract{We investigate how the left-hand cut (LHC) problem is treated in the HAL QCD method.
For this purpose, we first consider the effect of the LHC to the scattering problem in non-relativistic quantum mechanics with potentials. 
We show that
the $S$-matrix or the scattering phase shift obtained from the potential including the Yukawa term ($e^{- m_\pi r}/r$)
with the infra-red (IR) cutoff $R$
is well-defined even for the complex momentum $k$ as long as $R$ is finite,
and they are compared with those obtained by the analytic continuation without the IR cutoff.
In the $R\to\infty$ limit, the phase shift approaches the result from the analytic continuation at ${\rm Im}\, k < m_\pi/2$, 
while they differ at  ${\rm Im}\, k > m_\pi/2$,  
except $k= k_b$, where $k_b$ is the binding momentum.
We also observe that $k_b$ can be correctly obtained even at finite but large $R$.
Using knowledge obtained in the non-relativistic quantum mechanics, we present how we should treat the LHC in the HAL QCD potential method.
}

\FullConference{The 41st International Symposium on Lattice Field Theory (LATTICE2024)\\
 28 July - 3 August 2024\\
Liverpool, UK\\}

\begin{document}
\begin{flushright}
YITP-25-10, RIKEN-iTHEMS-Report-25
\end{flushright}
\maketitle

\section{Introduction}
The LHCb collaboration reported an experimental observation of a doubly charmed tetraquark state $T_{cc}^+$\cite{LHCb:2021vvq,LHCb:2021auc},
which appears around 360 keV below $D^{*+}D^0$ threshold as a narrow peak with $I(J^P)=0(1^+)$ in the $D^0D^0\pi^+$ invariant mass spectrum, and thus is supposed to contain two charm quarks and two light anti-quarks. Therefore  $T_{cc}^+$ is a genuine tetraquark state, since it cannot mix with an ordinary meson made of quark and anti-quark due to a conservation of a charm number.  

\begin{figure*}[htb]
\centering
\includegraphics[height=5cm]{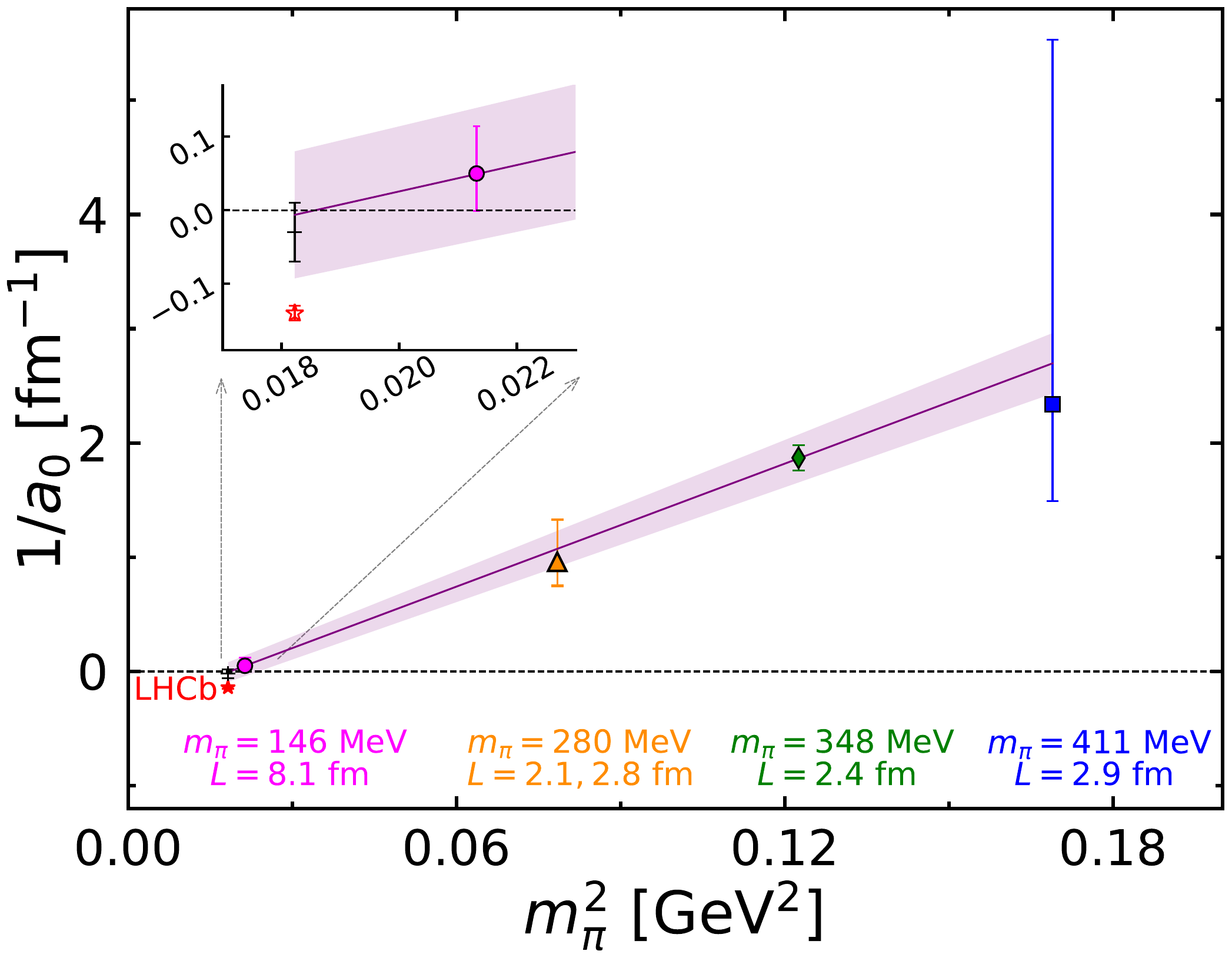}
\includegraphics[height=5cm]{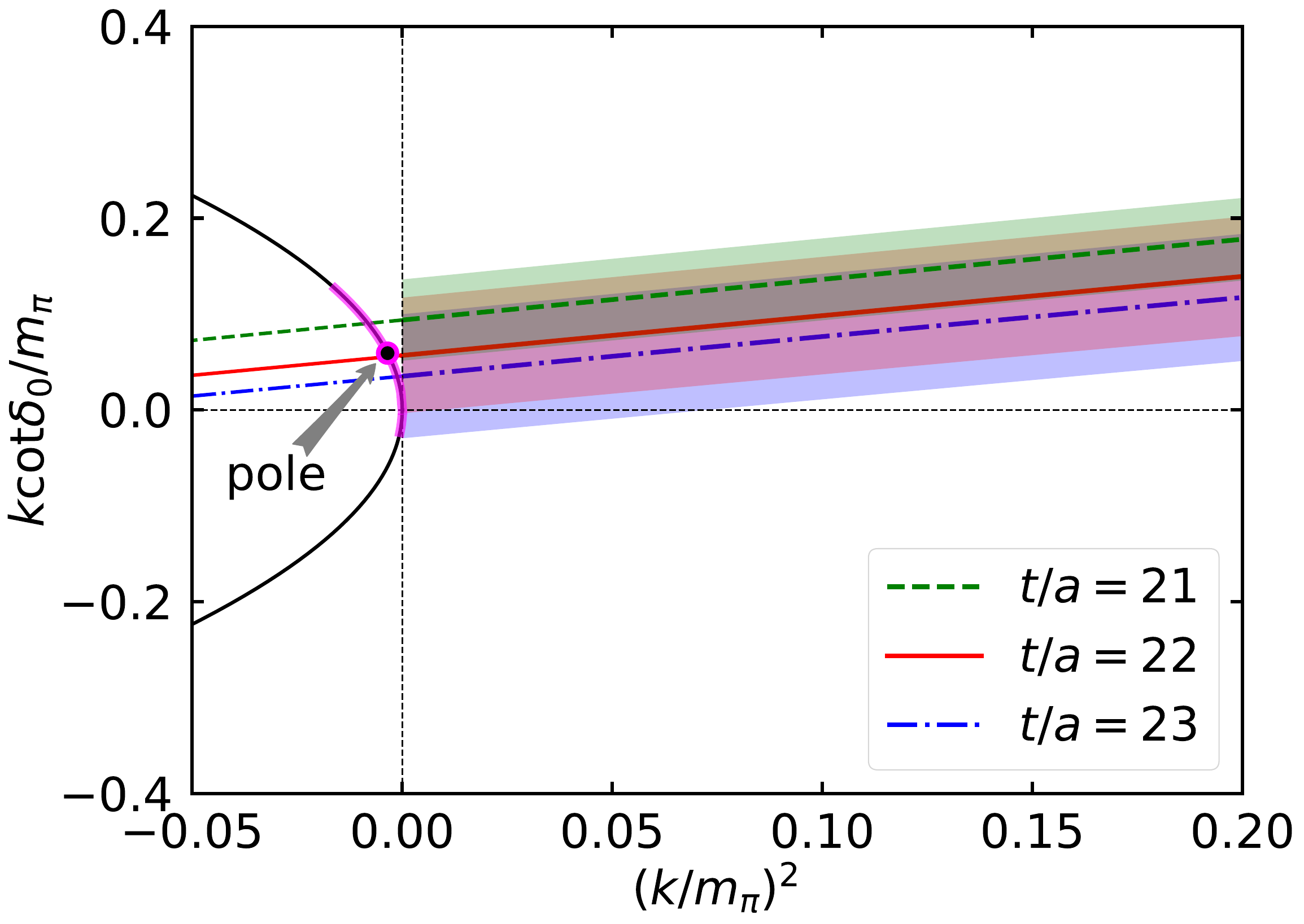}
\caption{(Left) The inverse scattering length $1/a_0$ for the $S$-wave $D^*D$ system with $I=0$ as a function of $m_\pi^2$, obtained from lattice QCD calculations by Refs.~\cite{Ikeda:2013vwa} (blue square),  \cite{Chen:2022vpo}(green diamond), \cite{Padmanath:2022cvl}(yellow triangle), and \cite{Lyu:2023xro}(magenta circle), together with its chiral extrapolation of $1/a_0$ linear in $m_\pi^2$ (violet solid line).
 The black plus shows the value obtained from the potential extrapolated to $m_\pi=135$ MeV\cite{Lyu:2023xro}, while
 the red star is the real part of the experimental value by LHCb\cite{LHCb:2021auc}.
 (Right) The $k\cot \delta_0/m_\pi$ as a function of $(k/m_\pi)^2$, where $\delta_0$ is the $S$-wave phase shift of the $D^*D$ scattering in the $I=0$.
 Its intersection with $\sqrt{ -(k/m_\pi)^2}$ (the upper black solid line) gives the position of the virtual pole.
 Both figures are taken from \cite{Lyu:2023xro}.
  }
 \label{fig:a0}
\end{figure*}

Studies on the $T^+_{cc}$ state in lattice QCD is summarized  in Fig.~\ref{fig:a0}(Left), where the inverse scattering length $1/a_0$ 
for the $S$-wave $D^*D$ system in the $I=0$ channel is plotted as a function of $m_\pi^2$.
These data are extrapolated  to $m_\pi =135$ MeV linearly in $m_\pi^2$, leading to $1/a_0$ = -0.01(9)  [fm$^{-1}$], which
is consistent with $1/a_0$ = -0.03(4)  [fm$^{-1}$] (black plus), obtained from the potential extrapolated to $m_\pi=135$ MeV\cite{Lyu:2023xro}. 
Thus  $T^+_{cc}$ seems to appear as a very shallow (quasi) bound state of $D^*D$ at the physical pion mass.

However, it has been pointed out that the analysis in some lattice studies to obtain the scattering length in Fig.~\ref{fig:a0} may not be justified due to the presence of the left-hand cut (LHC)\cite{Du:2023hlu}. 
For example, in Ref.~\cite{Padmanath:2022cvl}, the effective range expansion (ERE) was employed to obtain  the scattering length from the finite volume energies, some of which, however, are out of the range for the ERE to be valid.
To overcome difficulties associated with the LHC, analyses based on effective theories\cite{Du:2023hlu,Meng:2023bmz,Collins:2024sfi} as well as a modified finite volume formula in the presence of the LHC\cite{Raposo:2023oru} have been proposed.

Does the LHC also affect the result in Ref.~\cite{Lyu:2023xro} using the HAL QCD method ?
Fortunately, the virtual pole obtained in Ref.~\cite{Lyu:2023xro} appears above the branch point of the LHC, as shown in Fig.~\ref{fig:a0} (Right),
where the LHC appears below $(k/m_\pi)^2\simeq -0.02$. Therefore  the analysis in  \cite{Lyu:2023xro} is unaffected by the LHC and thus valid.
We however think that we need to understand what happens if a bound state appears on the LHC in the HAL QCD method for future studies. 

\section{Bound state on LHC}
\subsection{Left-hand cut}
We show how the LHC appears in the  $D^*D$ scattering.
The virtual pion propagator in the $u$ channel scattering in the center of mass reads
\begin{equation}
\frac{1}{ ( E_{D^*} - E_{D} )^2 -(\vec q-\vec p)^2-m_\pi^2}
=\frac{1} {m_{D^*}^2+m_D^2 -m_\pi^2 -2 E_{D^*} E_D + 2 k^2 \cos\theta},
\end{equation}
where $E_{D^*} =\sqrt{m_{D^*}^2+k^2}$ and  $E_{D} =\sqrt{m_{D}^2+k^2}$ with $k:=\vert \vec q\vert =\vert \vec p\vert$,
and $\vec q\cdot \vec p = k^2\cos\theta$. After the $S$ wave projection, the $u$ channel contribution is proportional to
 \begin{equation}
 \propto {1\over k^2}\log
\left[
\frac{m_{D^*}^2+m_D^2-m_\pi^2-2 E_{D^*} E_D + 2k^2}{m_{D^*}^2+m_D^2-m_\pi^2-2 E_{D^*} E_D - 2k^2}
\right],
\end{equation}
which generates two branch points at
\begin{equation}
k_\pm^2 = \frac{[(m_{D^*}-m_D)^2 -m_\pi^2][(m_{D^*}+m_D)^2 -m_\pi^2]}{4 A_\pm}, 
\quad  A_+= m_\pi^2 \ll A_- = 2( m^{2}_{D^*}+m^{2}_D) -m_\pi^2 .
\end{equation}
Therefore, if $ (m_{D^*}-m_D)^2 -m_\pi^2 < 0 $, there appears a cut at negative $k^2$, that is the left-hand cut.  

\subsection{LHC in the potential}
In this subsection, we discuss that the LHC problem appears also in quantum mechanics.
For the spherically symmetric potential, the $S$-wave Schr\"odinger equation reduces to
\begin{equation}
\left[{d^2\over dr^2} - U(r) + k^2 \right] \varphi(k,r) = 0,
\end{equation}
where $k$ is related to the energy as $k^2= 2M E$. For example, in the case of the Yukawa potential, we have
\begin{equation}
U(r) = g {e^{-m_\pi r}\over r},
\end{equation}
which mimics the non--relativistic contribution of the virtual pion exchange with its mass $m_\pi$. 

The $S$-matrix is defined as $S(k) = a/b$, 
where $a,b$ are determined from the asymptotic behavior of the regular solution satisfying $\varphi(k,0)=0$ and $\dfrac{d}{dr}\varphi(k,0) = 1$
as 
\begin{equation}
\varphi (k,r) \stackrel{r\to\infty}{\longrightarrow} ae^{i k r} - be^{-ikr} .
\end{equation}
The regular solution is explicitly given by
\begin{equation}
\varphi(k,r) ={1\over 2i k} \left[ e^{ikr} {\cal F}(-k,r) - e^{-ikr} {\cal F}(k,r)\right], \quad
{\cal F}(k,r)= 1+\int_0^r dr^\prime  e^{i kr^\prime} U(r^\prime) \varphi(k,r^\prime) . 
\end{equation}
Therefore, if $U(r)$ has the IR cutoff $R$ as $U(r > R) = 0 $, $S(k)$ is well-defined for $^\forall k \in \mathbb{C}$ as
\begin{equation}
S(k) =\frac{{\cal F}(-k,\infty)} {{\cal F}(k,\infty)} = \frac{{\cal F}(-k,R)} {{\cal F}(k,R)}:=S(k,R),
\end{equation}
so that there appears no LHC in $S(k,R)$. Since $U(r> R) =0$ is physically a good approximation for $R \gg 1/m$, the LHC problem  is rather
academic. In other words, we may say who cares the  tail of the potential behind the moon.
Nonetheless, for the academic nature of the problem, we investigate how $S(k,R)$  in the large $R$ limit is different from an analytic continuation of $S_{\rm anal}(k)$, obtained by the analytic continuation at $k^2\ge 0$ of the $S$-matrix  without IR cutoff.

\subsection{Large $R$ limit}
In the upper $k$ plane (${\rm Im}\ k \ge 0$ ) , ${\cal F}(k,R)$ is convergent as $R\to\infty$, while 
\begin{eqnarray}
{\cal F}(-k,R) &=& 1 +\int_0^R dr^\prime\, e^{-i k r^\prime} U(r^\prime) \varphi(k,r^\prime) ={\cal F}(-k,\bar R)
+\int_{\bar R}^R dr^\prime\, e^{-i k r^\prime} U(r^\prime) \varphi(k,r^\prime)\nonumber \\
&\simeq& {\cal F}(-k,\bar R) - g\frac{{\cal F}(k,\infty)}{2ik} \int_{\bar R}^R dr^\prime\, e^{-i 2k r^\prime}\frac{e^{-m_\pi r^\prime}}{r^\prime},
\label{eq:FkR}
\end{eqnarray}
as $R\to\infty$, 
where we use
\begin{equation}
\varphi(k,r)\simeq -\frac{{\cal F}(k,\infty)}{2i k} e^{-ik r}
\end{equation}
at large $r$ for  ${\rm Im}\ k \ge 0$.
Since the integrand in the second term of Eq. \eqref{eq:FkR} contains $e^{(2{\rm Im} k - m_\pi) r^\prime}$,
${\cal F}(-k,R)$ is divergent (convergent)  for ${\rm Im} k > m_\pi/2$ ( ${\rm Im} k < m_\pi /2$ ) as $R\to \infty$.

If there appears a bound state at $k=k_b$ for ${\rm Im}\, k_b > 0$ as a pole of the $S$-matrix $S(k) = {\cal F}(-k,\infty)/{\cal F}(k,\infty)$,
it implies ${\cal F}(k,\infty)=0$. Therefore, according to   \eqref{eq:FkR}, ${\cal F}(-k_b,R)$ is convergent even if ${\rm Im}\, k_b > m_\pi /2$.
Furthermore, at finite $R$, $S(k,R)$ has a pole at $k= k_b(R)$, which is $R$-dependent in principle, and it converges to $k_b$ as 
$\displaystyle \lim_{R\to\infty} k_b(R) = k_b$. Thus, $S(k_b(R),R)$ diverges even at finite $R$.

In summary, in the $R\to\infty$ limit, $S(k,R)$ is convergent at  ${\rm Im}\, k < m_\pi /2$, while 
$S(k,R)$ is divergent at  ${\rm Im}\, k > m_\pi /2$ due to the divergence of ${\cal F}(-k,\infty)$ except $k=k_b$.
Moreover, $S(k_b(R),R)$ has a pole due to ${\cal F}(k_b(R),R)=0$ event at finite $R$, and the pole converges to $k_b$ as $R\to\infty$.

\subsection{Example: Yukawa plus Gaussian at short distance}
As an explicit example, we consider a potential containing the Yukawa term (long range) plus a Gaussian term  (short range).  
Parameters in the potential are adjusted so that one bound state appears below the branch point of the LHC. The scattering phase shift is calculated for $k^2 >0$, and it is analytically continued to $k^2 <0$ region. 
We also introduce the IR cutoff $R$ to the Yukawa part of the potential, and calculate the $S$-wave scattering phase shift $\delta_R(k)$  at  $k^2<0$,
using
\begin{equation}
k \cot \delta_R(k) = i k\frac{S(k,R) + 1}{S(k,R) -1}, \quad S(k,R) =\frac{{\cal F}(-k,R)}{{\cal F}(k,R)},
\label{eq:kcotd}
\end{equation}
where both ${\cal F}(k,R)$ and  ${\cal F}(-k,R)$ are convergent for the finite $R$. 
 
\begin{figure*}[htb]
\centering
  \includegraphics[height=5cm]{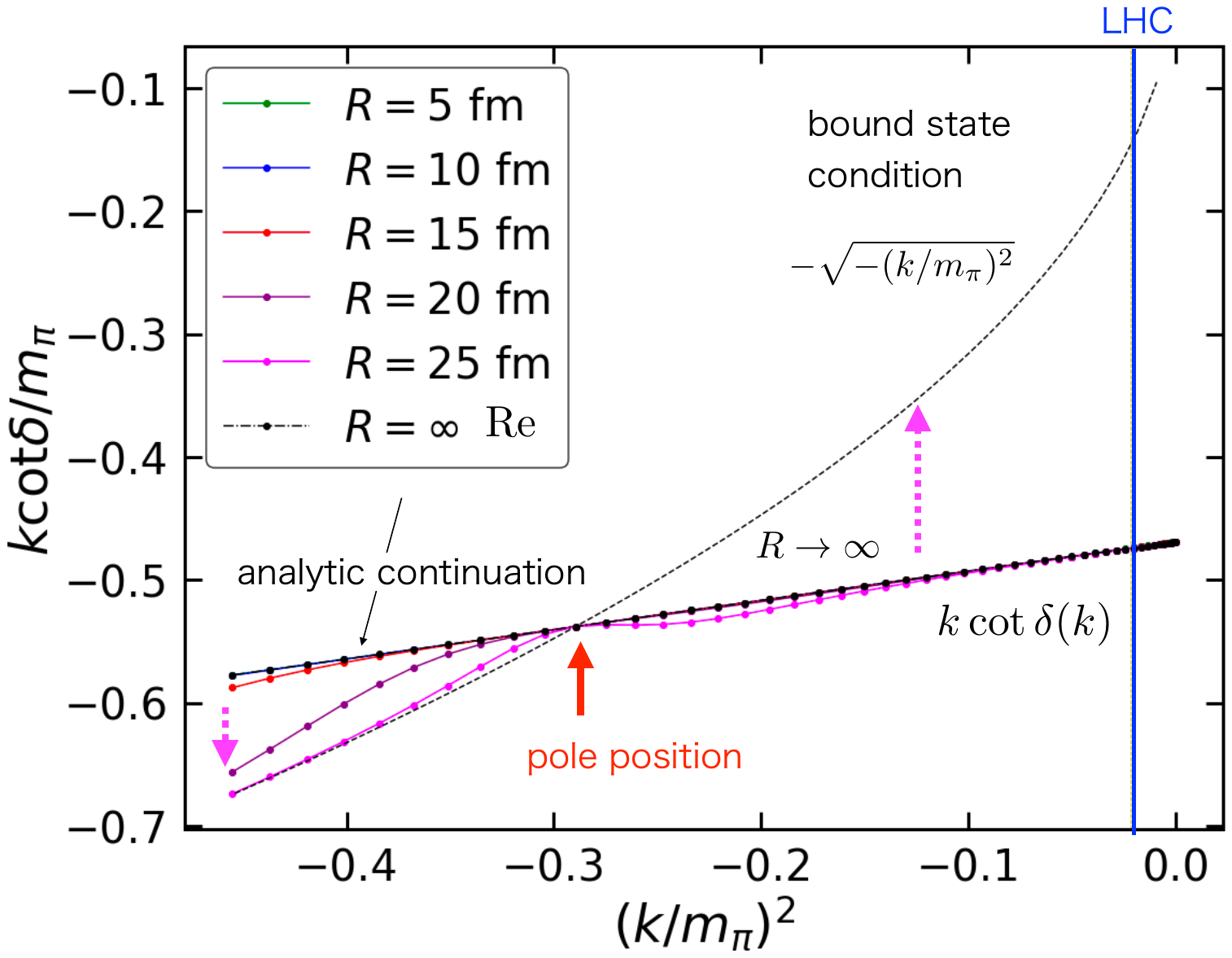}
    \includegraphics[height=5cm]{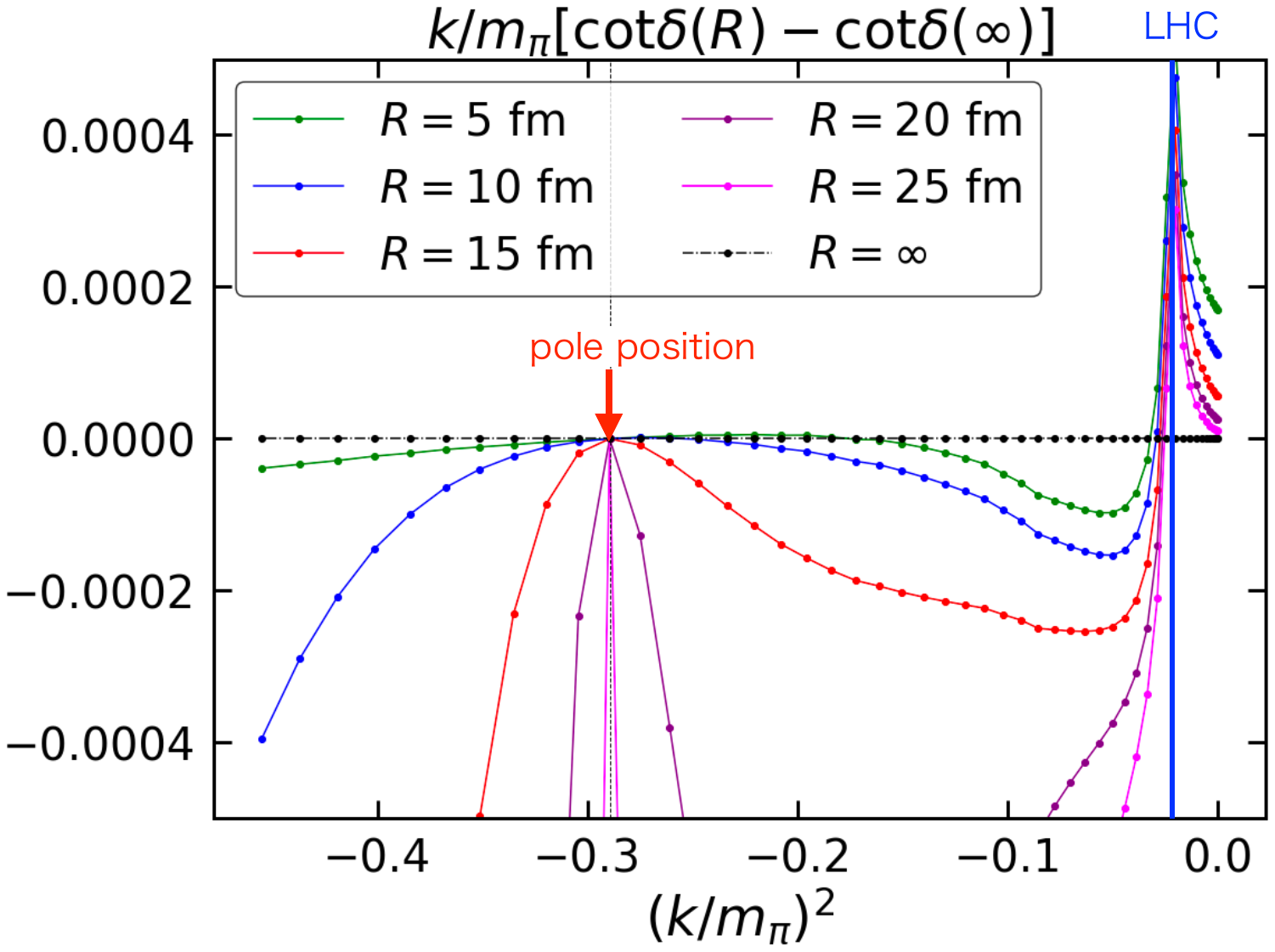}
 \caption{(Left) A comparison of  $ k\cot\delta(k)/m_\pi$ at $k^2<0$ region between the analytic continuation (black line, the real part only) and
 results with the IR cutoff, $R=$ 5 fm(green), 10 fm(blue), 15 fm(red), 20 fm(violet) and 25 fm(magenta), together with the bound state condition
 $ -\sqrt{-(k/m_\pi)^2}$ (black dashed line). 
(Right) A difference between  $ k\cot\delta_R(k)/m_\pi$ and the real part of $ k\cot\delta(k)/m_\pi$. 
  }
 \label{fig:ERE}
\end{figure*}

 We compare $k \cot \delta(k)$ obtained from the analytic continuation $S_{\rm anal}(k)$ with the result obtained from Eq. \eqref{eq:kcotd} for various values of $R$.
 In Fig.~\ref{fig:ERE} (Left), $k \cot \delta(k)/m_\pi$ from the analytic continuation is plotted by the black symbols, while
   $k \cot \delta_R(k)/m_\pi$  obtained from Eq. \eqref{eq:kcotd} are given for $R=$  5 fm (green), 10 fm (blue), 15 fm (red), 20 fm (violet) and 25 fm (magenta), 
   where we take $m_\pi= 146$ MeV.
These EREs intersect    with the bound state condition $-\sqrt{ -(k/m_\pi)^2}$ (black dashed line) at $k_b=k_b(\infty)$ and $k_b(R)$.
It is noted that $R$ dependence of $k_b(R)$ is very very small in this range of $R$, so that  we regard $k_b= k_b(R)$ hereafter.

 When ${\rm Im}\, k < m_\pi/2$, 
 $k\cot\delta_R(k)/m_\pi$  converges to $k\cot\delta(k)/m_\pi$, which is real in  this range of $k$. 
At ${\rm Im}\, k > m_\pi/2$,  while $k\cot\delta_R(k)/m_\pi$ at $R=5$ fm agrees with the real part of  $k\cot\delta(k)/m_\pi$,
it gradually deviates from it as $R$ increases.
Since $S(k,R)$ diverges as $R\to \infty$,  $k\cot\delta_R(k)/m_\pi$  should approach to the bound state condition $-\sqrt{-(k/m_\pi)^2}$ (black dashed line) as $R\to \infty$. Such behaviors observed in the left of the pole position, in particular, at $R = 25$ fm (magenta),
while   $k\cot\delta_R(k)/m_\pi$ is still close to the real part of  $k\cot\delta(k)/m_\pi$ in the right of the pole position.
This difference can be understood since the divergence is characterized by a factor $e^{(2{\rm Im} k - m_\pi) r^\prime}$ in the integrand of Eq. \eqref{eq:FkR}.
The larger ${\rm Im}\, k$ is, the stronger the divergence factor becomes as $R$ increases.

Since $k_b(R)$ is almost $R$-independent in this range of $R$, we can easily estimate the correct value of the binding momentum $k_b$ from the
$k\cot\delta_R(k)$ at finite but large $R$ without the $R\to\infty$ limit, even though $k\cot\delta_R(k)/m_\pi$ is different from $k\cot\delta(k)/m_\pi$
at  ${\rm Im}\, k > m_\pi/2$ other than  $k_b$.   
While $S(k,R)$ is always real for pure imaginary $k$, $S_{\rm anal}(k)$ is complex for ${\rm Im}\, k > m_\pi/2$.

In Fig.~\ref{fig:ERE} (Right), a difference, $k\cot \delta_R(k) - {\rm Re}\, [k\cot \delta/m_\pi]$ are plotted.
Properties mentioned above are more clearly seen.
(1) $S(k,R)$ converges to $S_{\rm anal}(k)$ at ${\rm Im}\, k< m_\pi/2$ as $R$ increases.
(2) $S(k,R)$ deviates from $S_{\rm anal}(k)$ at ${\rm Im}\, k > m_\pi/2$ except $k=k_b$.
(3) The pole position of $S(k,R)$ reproduces the binding momentum $k_b$ correctly.

\section{HAL QCD method with LHC}
Based on knowledge obtained in the previous section, we discuss how we can treat the LHC in the HAL QCD method.
Below we summarize the procedure to deal with the LHC in the HAL QCD method.

\begin{enumerate}
\item Since the potential is obtained in the HAL QCD method, we first investigate the long distance behavior of the potential.
For example, the $D^*D$ potential in the $I=0$ and $S$-wave channel, which is relevant for $T^+_{cc}$, is shown in Fig.~\ref{fig:potential} (Left).
The 2-pion  rather than 1-pion  exchange has been observed at the long distance of the potential\cite{Lyu:2023xro}.
Thus the potential is fitted by 2-Gaussians + Yukawa$^2$,
\begin{equation}
V_{\rm fit}^{2\pi}(r;m_\pi) =\sum_{i=1}^2 a_i e^{-r^2/b_i^2} + a_3 \left(1 - e^{-r^2/b_3^2} \right)^2\left(\frac{e^{-m_\pi r}}{r}\right)^2,
\end{equation}
as shown in the red band in the figure.
Note that the 1-pion exchange, which is expected to exist, is not a dominant contribution in the  $D^*D$ potential for $T^+_{cc}$.

\begin{figure}[htb]
\centering
  \includegraphics[angle=0, width=0.43\textwidth]{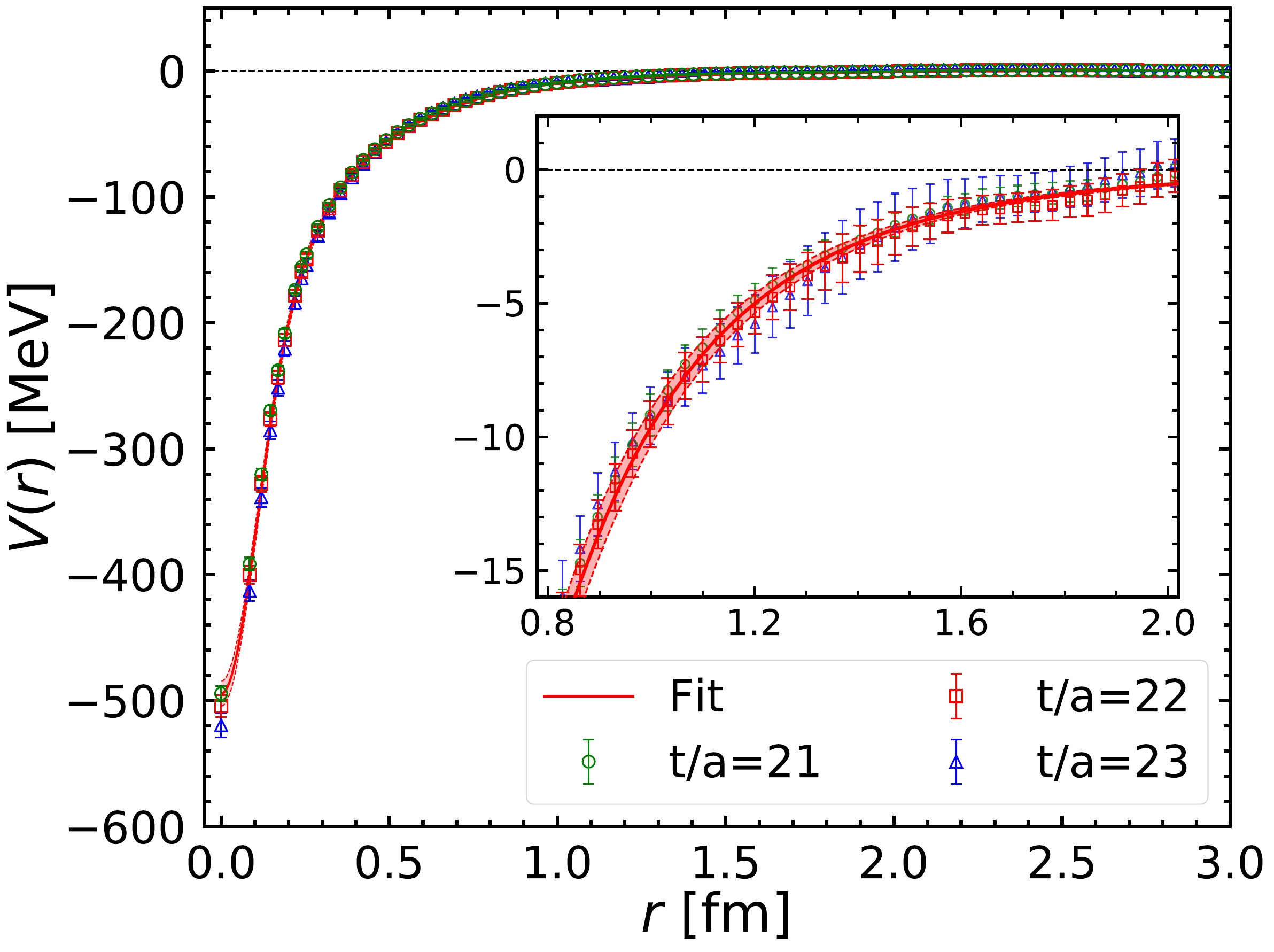}
  \hskip 0.5cm
    \includegraphics[angle=0, width=0.53\textwidth]{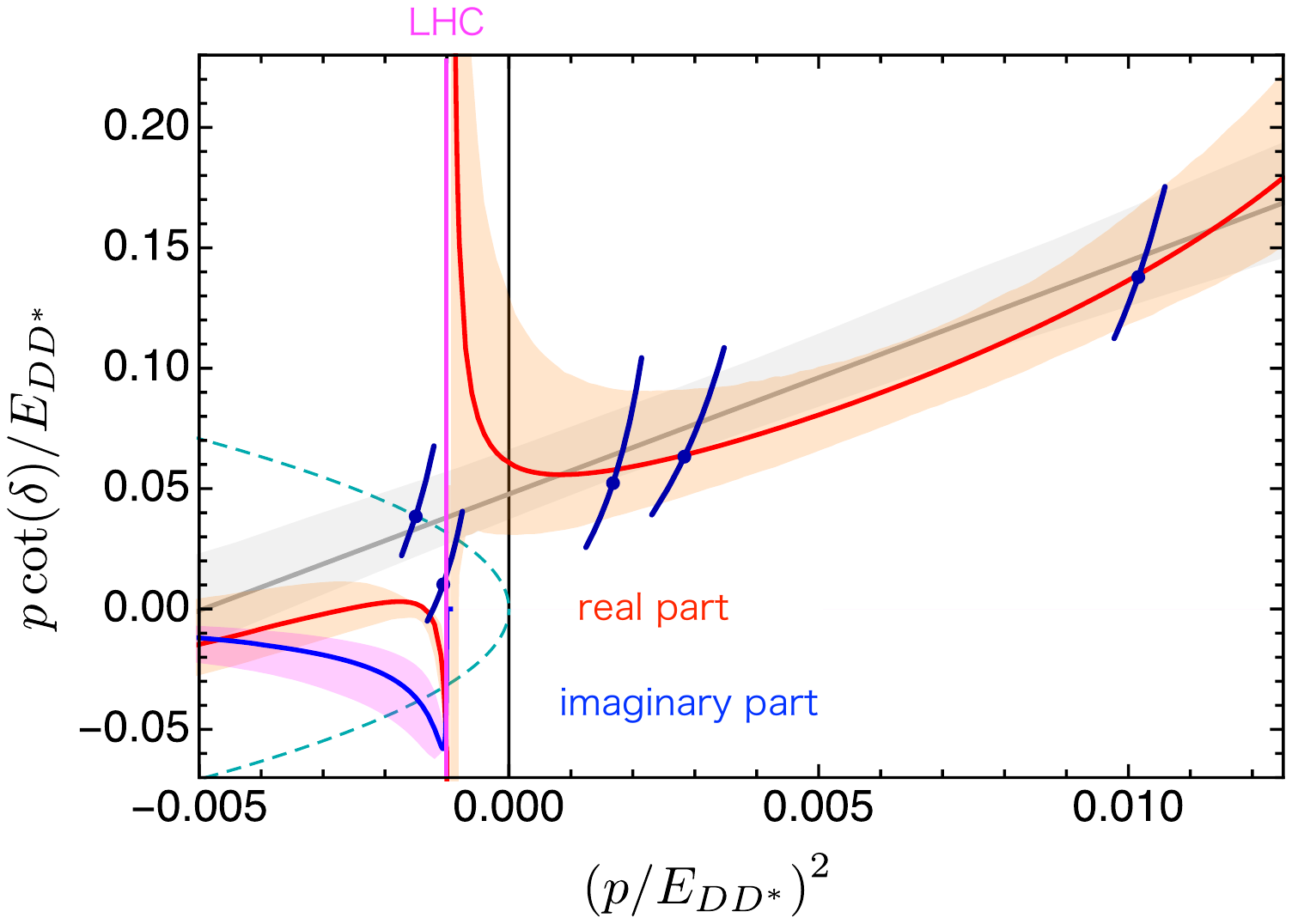}
 \caption{(Left) The $D^* D$ potential $V(r)$ in the $I=0$ and $S$-wave channel as a function of $r$, obtained in lattice QCD
 at Euclidean time $t/a =21$ (green circles), 22 (red squares) and 23 (blue triangles), 
 together with the fit $V_{\rm fit}^{2\pi}(r;m_\pi)$ to data at $t/a=22$ (red band). This figure is taken from \cite{Lyu:2023xro}.
 (Right) The reanalysis of $p\cot\delta(p)/E_{DD^*}$ data in \cite{Padmanath:2022cvl} by including the effect of the LHC\cite{Du:2023hlu},
 where red and blue lines represent real and imaginary parts of  $p\cot\delta(p)/E_{DD^*}$.
 We added few words in the figure taken from \cite{Du:2023hlu}.
  }
 \label{fig:potential}
\end{figure}

\item Next, we estimate positions of the LHC. For example, the LHC for the 2-pion exchange appears at
\begin{equation}
\frac{k_{-}^2}{m_\pi^2}\simeq \frac{ (m_{D^*}-m_D)^2-(2m_\pi)^2}{4m_\pi^2} \simeq -0.77,
\end{equation}
while for the 1-pion exchange, it becomes
\begin{equation}
\frac{k_{-}^2}{m_\pi^2}\simeq \frac{ (m_{D^*}-m_D)^2-(m_\pi)^2}{4m_\pi^2} \simeq -0.02,
\end{equation}
where $m_\pi=146.4$ MeV, $m_D=1878.2$ MeV and $m_{D^*}=2018.1$ MeV in the simulation\cite{Lyu:2023xro} are used.

\item We then compare the position of the LHC with the analytic continuation of $k\cot\delta(k)$.
If a bound or virtual state appears above the branch point of the LHC, as seen in Fig.~\ref{fig:a0} (Left),  the analysis by the analytic continuation is valid.
If a bound state appears below the branch point of
 the LHC, we should directly extract the binding energy by solving the Schr\"odiner equation,
rather than finding a crossing point between  $k\cot\delta(k)$ and the bound state condition $-\sqrt{-k^2}$.

\item If a virtual state appears below the branch point of the LHC, the required analysis 
depends on the method to obtain data.
In the case of the finite volume method,
Fig.~\ref{fig:potential} (Right) shows 
an example of the analysis 
including the LHC\cite{Du:2023hlu}, applied to finite volume spectra\cite{Padmanath:2022cvl}. On the LHC, both real (red) and imaginary (blue) parts of $p\cot\delta(p)/E_{DD^*}$ appear, where $E_{DD^*}=\sqrt{p^2+m_D^2}+\sqrt{p^2+ m_{D^*}^2}$ , so that virtual pole on real $k^2$ claimed in \cite{Padmanath:2022cvl} disappears.
In the case of the HAL potential, however, the situation is different.
Since the long distance behavior of the potential controls the position of the LHC, as discussed in the previous section,
the analytic continuation knows the existence of the LHC, so that $k\cot\delta(k)$ automatically becomes complex on the LHC.
Therefore, for example, in the case of $V^{2\pi}_{\rm fit}(r;m_\pi)$ including the 2-pion exchange, either virtual pole appears above the LHC of the 2-pion
exchange at $k^2 \simeq - 0.77 m_\pi^2$ or no virtual pole exists.

\item If the LHC is expected to exist, we may include it in the potential fit.
For example, even though the 1-pion exchange contribution is not seen in the $D^*D$ potential,
we may include it as an alternative fit to estimate its effects as
\begin{equation}
V^{+1\pi}_{\rm fit}(r;m_\pi)=V^{2\pi}_{\rm fit}(r;m_\pi) + a_4 \left(1 - e^{-r^2/b_4^2} \right) \frac{e^{-m_\pi r}}{r}.
\end{equation}  
 The analysis with the alternative fit gives an estimate for systematic errors.

Note that the binding energy is not affected by the LHC, if the Schr\"odinger equation is directly solve.
A difference between binding energies from $V^{2\pi}_{\rm fit}(r;m_\pi)$ and $V^{+1\pi}_{\rm fit}(r;m_\pi)$, if exists,  is regarded as a systematic error, too.

\end{enumerate}

\section{Conclusion and discussion}
The analysis using the non-relativistic potential shows that
the analytic continuation and the result in the limit of the infra-red cutoff ($R\to \infty$) differ below the branch point of the LHC, except the binding momentum $k_b$.
Thus the binding energy in the HAL QCD method is not affected by the LHC even if it exists.
For the virtual state below the LHC, the analytic continuation can be performed through the HAL QCD potential in order to include an effect of the LHC.
The long distance behavior which causes a particular type of the LHC may be included in the fit of the HAL QCD method, even if data show no such behavior.
A doubly charmed tetra-quark state $T_{cc}^+$ in the previous HAL QCD method\cite{Lyu:2023xro} appears as a virtual state above the branch point of the 1-pion LHC, and
thus remains valid.

In order control the effect of the LHC, it is essential to determine the long distance behavior of the potential. Therefore the HAL QCD method is a more suitable framework than the finite volume method to do this explicitly.

\section*{Acknowledgment}
This work has been supported in part by  
the JSPS (Grant Nos. JP23H05439, JP22H00129 and JP19K03879), ``Program for Promoting Researches on the Supercomputer Fugaku'' (Simulation for basic science: from fundamental laws of particles to creation of nuclei)
and (Simulation for basic science: approaching the new quantum era) (Grant Nos. JPMXP1020200105 and  JPMXP1020230411).


\begin{thebibliography}{99}

\bibitem{LHCb:2021vvq}
R.~Aaij \textit{et al.} [LHCb],
Nature Phys. \textbf{18} (2022) no.7, 751-754
doi:10.1038/s41567-022-01614-y
[arXiv:2109.01038 [hep-ex]].

\bibitem{LHCb:2021auc}
R.~Aaij \textit{et al.} [LHCb],
Nature Commun. \textbf{13} (2022) no.1, 3351
doi:10.1038/s41467-022-30206-w
[arXiv:2109.01056 [hep-ex]].

\bibitem{Ikeda:2013vwa}
Y.~Ikeda, B.~Charron, S.~Aoki, T.~Doi, T.~Hatsuda, T.~Inoue, N.~Ishii, K.~Murano, H.~Nemura and K.~Sasaki,
Phys. Lett. B \textbf{729} (2014), 85-90
doi:10.1016/j.physletb.2014.01.002
[arXiv:1311.6214 [hep-lat]].

\bibitem{Chen:2022vpo}
S.~Chen, C.~Shi, Y.~Chen, M.~Gong, Z.~Liu, W.~Sun and R.~Zhang,
Phys. Lett. B \textbf{833} (2022), 137391
doi:10.1016/j.physletb.2022.137391
[arXiv:2206.06185 [hep-lat]].

\bibitem{Padmanath:2022cvl}
M.~Padmanath and S.~Prelovsek,
Phys. Rev. Lett. \textbf{129} (2022) no.3, 032002
doi:10.1103/PhysRevLett.129.032002
[arXiv:2202.10110 [hep-lat]].

\bibitem{Lyu:2023xro}
Y.~Lyu, S.~Aoki, T.~Doi, T.~Hatsuda, Y.~Ikeda and J.~Meng,
Phys. Rev. Lett. \textbf{131} (2023) no.16, 161901
doi:10.1103/PhysRevLett.131.161901
[arXiv:2302.04505 [hep-lat]].

\bibitem{Du:2023hlu}
M.~L.~Du, A.~Filin, V.~Baru, X.~K.~Dong, E.~Epelbaum, F.~K.~Guo, C.~Hanhart, A.~Nefediev, J.~Nieves and Q.~Wang,
Phys. Rev. Lett. \textbf{131} (2023) no.13, 131903
doi:10.1103/PhysRevLett.131.131903
[arXiv:2303.09441 [hep-ph]].

\bibitem{Meng:2023bmz}
L.~Meng, V.~Baru, E.~Epelbaum, A.~A.~Filin and A.~M.~Gasparyan,
Phys. Rev. D \textbf{109} (2024) no.7, L071506
doi:10.1103/PhysRevD.109.L071506
[arXiv:2312.01930 [hep-lat]].

\bibitem{Collins:2024sfi}
S.~Collins, A.~Nefediev, M.~Padmanath and S.~Prelovsek,
Phys. Rev. D \textbf{109} (2024) no.9, 9
doi:10.1103/PhysRevD.109.094509
[arXiv:2402.14715 [hep-lat]].

\bibitem{Raposo:2023oru}
A.~B.~Raposo and M.~T.~Hansen,
JHEP \textbf{08} (2024), 075
doi:10.1007/JHEP08(2024)075
[arXiv:2311.18793 [hep-lat]].

\end{thebibliography}
\end{document}